\documentclass[aps, twocolumn, pre, floatfix, superscriptaddress]{revtex4-2}

\usepackage{graphicx}
\usepackage{bm}
\usepackage{bbold}
\usepackage{amsmath,amssymb,amsfonts,amsthm}
\usepackage{physics}
\usepackage{color}

\usepackage{tabularx,dcolumn} 

\usepackage{ulem}

\definecolor{darkblue}{rgb}{0,0,.65}
\definecolor{darkgreen}{rgb}{0.3,0.6,0.3}
\definecolor{cyan1}{rgb}{0.0, 0.6, 0.6}
\usepackage[%
pdfstartview=FitH,%
breaklinks=true,%
bookmarks=true,%
colorlinks=true,%
anchorcolor=black,%
citecolor=blue,
filecolor=black,%
menucolor=black,%
urlcolor=darkblue,%
linkcolor=blue,%
]{hyperref}

\providecommand{\outerprod}[2]{\ensuremath{| #1 \rangle \langle #2 | }}
\providecommand{\ket}[1]{\ensuremath{\left|{#1}\right.\rangle}}
\providecommand{\bra}[1]{\ensuremath{\langle\left.{#1}\right|}}

\newtheorem*{conjecture}{Conjecture}

\graphicspath{{./Figures/}}

\begin{document}
	
	\title{Structure of the Hamiltonian of mean force}
	
	\author{Phillip C. Burke}
	
	\affiliation{School of Physics, University College Dublin, Belfield, Dublin 4, Ireland}
	\affiliation{Centre for Quantum Engineering, Science, and Technology, University College Dublin, Dublin 4, Ireland}
	\affiliation{Department of Theoretical Physics, Maynooth University, Maynooth, Kildare, W23 F2H6, Ireland}
	
	\author{Goran Nakerst}
	
	\affiliation{Institut f\"ur Theoretische Physik, Technische Universit\"at Dresden, 01062 Dresden, Germany}
	
	\author{Masudul Haque}
	
	\affiliation{Institut f\"ur Theoretische Physik, Technische Universit\"at Dresden, 01062 Dresden, Germany}
	
	
	
	\date{\today}
	
	\begin{abstract}
		The Hamiltonian of mean force is an effective Hamiltonian that allows a quantum system, non-weakly coupled to an environment, to be written in an effective Gibbs state. We present results on the structure of the Hamiltonian of mean force in extended quantum systems with local interactions. We show that its spatial structure exhibits a ``skin effect'' --- its difference from the system Hamiltonian dies off exponentially with distance from the system-environment boundary.  For spin systems, we identify the terms that can appear in the Hamiltonian of mean force at different orders in the inverse temperature.   
		
	\end{abstract}
	
	\maketitle
	
	
	\begin{figure}
		\centering
		\includegraphics[width=0.30\linewidth]{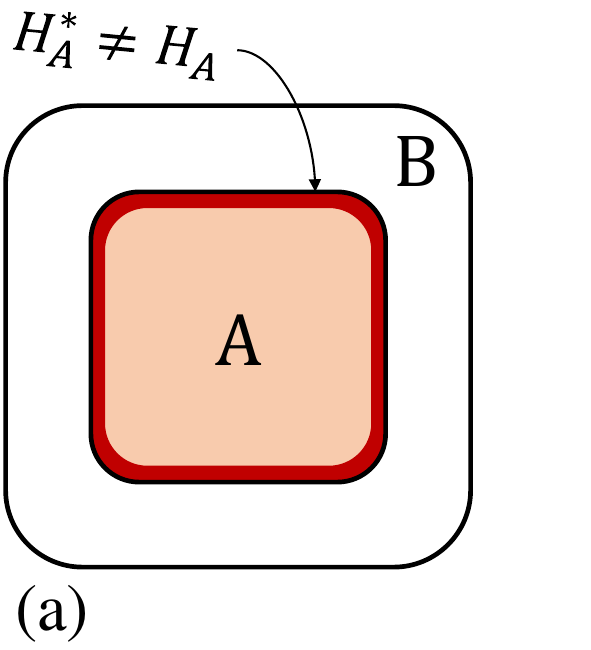}
		\includegraphics[width=0.68\linewidth]{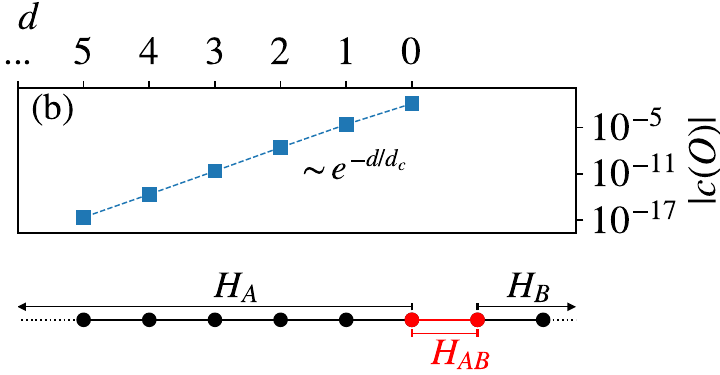}
		\caption{
			\textbf{(a)} Illustration of the skin effect in a composite system. The dark red region indicates the shallow effect of $B$ within $A$.
			\textbf{(b)} Magnitude of coefficients in the HMF plotted versus the distance $d$ from the boundary site $L_A$ at small $\beta$, illustrating the exponential decay with distance $d$.
		}
		\label{fig:1}
	\end{figure}
	
	
	\section{Introduction}
	
	The assumption of weak system-bath coupling is widespread in thermodynamics but is clearly not universally valid.
	There is thus considerable interest in formulating thermodynamics to include effects of non-weak coupling \cite{BinderGogolinAndersAdesso_book2018_ThermoInQuantumRegime}. 
	A central concept in this effort is that of the Hamiltonian of mean force (HMF) \cite{CampisiTalknerHanggi_HMF_PRL2009, Talkner_Hanggi_HMF_RevModPhys2020}, the quantum counterpart to the classical potential of mean force \cite{Kirkwood_PMF_1935, Jarzynski_StrongCoupledThermal_JoSM2004, Tuckerman_StatMech_2010}.   Consider a combined system in thermal equilibrium, composed of a system $A$ and a bath/environment $B$, interacting via the Hamiltonian $H_{AB}$ so that the combined Hamiltonian is $H=H_A+H_B+H_{AB}$. 
	The system state $\rho_A$, obtained by tracing out $B$ degrees of freedom, is then not a thermal Gibbs state in general and differs from $e^{-\beta H_{A}}$ 
	\cite{Gelin_Thoss_subensemble_PRE2009, Hilt_HMF_PRE2011, GallegoEisert_BeyondWeakCoup_NJoP2014}, here $\beta$ is the inverse temperature.  
	Nevertheless, we can formally define an effective Hamiltonian $H^{*}_{A}$ so that $\rho_A\propto e^{-\beta H^{*}_{A}}$ \cite{Rivas_StrongCoupOpenQu_PRL2020, CresserAnders_HMFcoupling_PRL2021, ChiuStrathearnKeeling_HMF_PRA2022, ChenCheng_HMF_JocP2022, Teretenkov_EffectiveGibbState_Ent2022, TrushechkinCresserAnders_MFGibbs_AVS2022, TimofeevTrushechkin_HMF_JoMPA2022, LeeYeo_SteadyStatesQME_PRE2022, AntoSztrikacsSegal_EffectHam_PRXQ2023}.   This $H^{*}_{A}$ is the HMF.  For non-negligible coupling $H_{AB}$, the HMF will generally deviate from the actual system Hamiltonian $H_A$. 
	For extended systems with local degrees of freedom, we address the question of how $H^{*}_{A}$ differs from $H_A$, in particular, the spatial dependence of this difference.  
	
	The HMF often appears implicitly in the study of thermalization in isolated quantum systems 
	\cite{AlessioRigol_Chaos_AdvPhys2016, NandkishoreHuse_MBL_AnnRev2015,Muller_ThermalizationTypicality_CiMP2015, Greinergroup_thermalization_Science2016,Dymarsky_Lashkari_Liu_PRE2018, GarrisonGrover_SingleEstate_PRX2018, Burke_Nakerst_Haque_temp_PRE2023}.
	Any spatial segment of an isolated system ($A$) can be regarded as being thermalized by the rest of the system ($B$), so that the expectation values of local observables having support in $A$ are given by their expectation values in the reduced thermal state. Since the partition between $A$ and $B$ is arbitrary in this setup, the coupling $H_{AB}$ is not small, and thus the thermal state to be used is $\rho_A \propto \tr_Be^{-\beta H}\propto e^{-\beta H^{*}_{A}}$ rather than just $e^{-\beta H_{A}}$.
	In this context, the temperature $1/\beta$ is determined using the canonical-ensemble correspondence between energy and temperature \cite{Rigol_thermalization_Nat2008, Rigol_thermalization_PRL2009,
		Rigol_Quenches_PRA2009,RigolSantos_Chaos_PRA10,	SantosRigol_Thermalization_PRE2010,Roux_quantumquenches_PRA2010, RigolSrednicki_Thermalization_2012,SantosPolkovnikovRigol_typicality_PRE2012, NeuenhahnMarquardt_thermalization_PRE2012,RigolSrednicki_FDT_PRL2013,
		SorgVidmarHeidrichMeisner_thermalization_PRA14,NandkishoreHuse_MBL_AnnRev2015,
		FratusSrednicki_EstateThermalization_PRE2015,Essler_Quench_JoSM2016,
		Greinergroup_thermalization_Science2016,AlessioRigol_Chaos_AdvPhys2016,
		GarrisonGrover_SingleEstate_PRX2018,LuGrover_RenyiEntropy_PRE2019,	SekiYunoki_thermal_PRR2020,Noh_ETH_PRE2021}. 
	The relevance of the HMF concept in this setup raises the question of the structure of the HMF in systems with local interactions. 
	Two questions immediately arise: 
	(1) What is the spatial structure of the HMF, i.e., of the deviation of $H^{*}_{A}$ from $H_{A}$?
	(2) What type of interactions are contained in the HMF?  
	
	We find that when $\beta$ is not very large, the difference of coefficients of HMF terms from the corresponding $H_A$ coefficients decay exponentially with distance $d$ to the boundary between $A$ and $B$, as illustrated in Fig.~\ref{fig:1}(b). 
	This result implies a ``skin effect'', illustrated in Fig.~\ref{fig:1}(a): The HMF effectively only deviates from $H_A$ near the boundary, i.e., the `bath' $B$ has a very shallow influence in $A$. 
	This is a general result, provided the total system Hamiltonian is made of local terms.
	For $\beta\to\infty$ (very low temperature), we show how the HMF is related to the ``entanglement Hamiltonian'' \cite{LiHaldane_EntanglementSpectrum_PRL2008, PeschelEisler_RDMEnt_JoPA2009, EislerPeschel_PropEntHam_JoSM2018, DalmonteEisler_EntHamRev_AndP2022} which is calculated from the ground state alone.
	We formulate our results in terms of spin chains, which are paradigm models for extended quantum systems with local interactions. 
	Using a perturbative framework coupled with numerical analysis, we elucidate the types of terms that can appear in the HMF and present systematics on which types of terms can appear at which order in $\beta$. 
	
	\section{Definitions \& Setup}
	%
	The combined system is taken to be in a thermal state, $\rho=e^{-\beta H}/Z$, so that subsystem $A$ (the `system') is in state $\rho_A = \frac{1}{Z}\tr_B e^{-\beta H}$,  where $\tr_B$ is the partial trace over $B$.   The HMF $H^{*}_{A}$ is defined to satisfy 
	\cite{CampisiTalknerHanggi_HMF_PRL2009},
	\begin{equation}\label{eq:hmf_def}
		\rho_A \ = \ \frac{1}{Z^*(\beta)} e^{-\beta H^{*}_{A}(\beta)}.
	\end{equation}
	
	This essentially defines the HMF only up to a constant \cite{StrasbergEsposito_MeasurabilityHMF_PRE2020, TalknerHangii_Comment_PRE2020, StrasbergEsposito_ReplyToComment_PRE2020}. As this freedom is not of interest to us (it only adds an identity operator to the HMF), we remove the ambiguity by choosing the normalization to be $Z^*(\beta) = Z(\beta)/Z_B(\beta)$, where  $Z$ and $Z_B$ are the partition functions of the combined system and the subsystem $B$ (`bath'), respectively.
	With this definition, the HMF is  
	\begin{equation}
		\label{eq:hmf_fulldef}
		H^{*}_{A}(\beta) = -\frac{1}{\beta} \ln \frac{\tr_B(e^{-\beta H})}{\tr_B(e^{-\beta H_B})}.
	\end{equation}
	The HMF $H^{*}_{A}$ reduces to the bare system Hamiltonian $H_A$ for vanishing coupling ($H_{AB}=0$) \cite{Goldstein_CanonicalTyp_PRL2006, Popescu2006, Mori_Ikeda_Ueda_thermalizationreview_JPB2018} and for infinite temperature ($\beta=0$) \cite{ChenCheng_HMF_JocP2022}.

	
	
	
	For definiteness, we formulate our results for the \textit{XXZ} chain, with and without magnetic fields. 
	This exemplifies the case of operators being supported on nearest neighbor sites (spin-spin interactions) or single sites (magnetic fields). 
	Generalization to longer-range interactions and to fermionic or bosonic systems should be straightforward, but we do not attempt to write out explicitly all such cases. 
	The Hamiltonian is 
	\begin{align}\label{eq:XXZ_chain}
		H \ = \ J \sum_{j=1}^{L-1}&(\sigma^x_j \sigma^x_{j+1} + \sigma^y_j \sigma^y_{j+1}) + \Delta \sum_{j=1}^{L-1} \sigma^z_j \sigma^z_{j+1} \notag \\
		& +\sum_{j=1}^{L} (h_j^z \sigma^z_j + h_j^x \sigma^x_j),
	\end{align}
	where $J$ is the site-to-site `hopping' strength, $\Delta$ is the spin-spin interaction strength (anisotropy), and $h_j^z$, ($h_j^x$) denotes the strength of the on-site magnetic field in the $z$ direction ($x$ direction) on-site $j$.
	Unless otherwise stated, we will generally use uniform magnetic fields, i.e., $h_j^z = h^z$ and $h_j^x = h^x$, $\forall j$. The case of $h_j^x = h_j^z = 0$ is the standard \textit{XXZ} chain.
	%
	We take the first $L_A$ sites as subsystem $A$; the boundary bond connects sites $j=L_A$ and $j=L_A+1$. 
	
	A convenient basis to investigate the HMF of spin systems is the basis of Pauli operators. The Pauli operators $O=\sigma_{i_1}^{\alpha_1}\dots\sigma_{i_{L_A}}^{\alpha_{L_A}}$, where $\alpha_j=\{x,y,z\}$ and $i_j$ denote sites in $A$, together with the identity operator form an orthogonal basis of operators with respect to the Hilbert-Schmidt scalar product. 
	The coefficient of an operator in a Hamiltonian is given by the scalar product of the operator with the Hamiltonian. 
	Thus, we quantify the deviation of the HMF from $H_A$ by
	\begin{equation}
		c(O) = \tr\left[O \cdot \left(H^{*}_{A}(\beta) - H_A\right)\right],
	\end{equation}
	i.e., by the difference of coefficients of the same operator in the HMF and $H_A$.   
	
	All numerical results in this work are obtained via exact diagonalization, i.e., the canonical density matrix of the entire $A+B$ system is constructed explicitly as a $2^L\times2^L$ matrix, and then the $B$ degrees of freedom are traced out numerically.  The computations are performed with 128-bit precision. 
	This higher precision is necessary because in many cases the coefficients are orders of magnitude smaller than the precision of double-precision floating-point arithmetic ($\sim 10^{-16}$).
	
	We refer to the $B$ partition (sites $L_A+1$ to $L$) as the `bath'.  However, in the present setting, we are not interested in the thermalizing effect of the bath, since the full $A+B$ system is already imposed to be in a Gibbs (thermal) state.  This means that the $B$ partition is not required to have any of the properties  (zero memory, fast timescales, infinite bandwidth, etc) that a physical bath is usually assumed to have.   In particular, the $B$ partition is not required to be larger than the $A$ partition.  In fact, we find that the size of the `bath' does not affect the qualitative insights presented in this work.  Therefore, it is computationally advantageous to take the $B$ partition to be as small as meaningful, and we present much of our data for systems with $L=L_A+1$, i.e., a single site in $B$.  This might appear to contradict the usual idea of a physical bath that thermalizes a system and sets the temperature, for which a large size is necessary or at least helpful.  However, in the present setting, a single-site `bath' is perfectly reasonable.

	\section{Which terms appear?}
	Terms appearing in $H^{*}_{A}$, beyond those already in $H_A$, are formed by combining terms in $H$, as can be seen by considering an expansion of Eq.~\eqref{eq:hmf_fulldef} in $\beta$. Thus any term in $H^{*}_{A}$ must have the form  $O \equiv h_1\cdots h_{k}$ (equality up to a constant), with the $h_i$ being terms that appear in $H$. 
	
	
	The type of Pauli operators appearing in $H$ constrains the types that can appear in $H^{*}_{A}$. 
	We outline the cases of an \textit{XXZ} chain -- other cases can be worked out analogously.
	For the \textit{XXZ} Hamiltonian \textit{without} magnetic fields, the operators appearing in the HMF can be identified by considering a homomorphism of the Klein four-group, the group of Pauli matrices modulo phases \cite{Suppmat}. The homomorphisms are given by assigning a sign $\pm 1$ to single-body Pauli operators $\sigma^x$, $\sigma^y$, and $\sigma^z$. All terms in the \textit{XXZ} Hamiltonian without magnetic fields have sign $+1$, regardless of the sign function. This carries over to the corresponding HMF \cite{Suppmat}. For single-body and mixed two-body ($\sigma_i^\alpha \sigma_j^{\alpha'}$ for $\alpha\neq\alpha'$) Pauli operators, there are sign functions such that the sign of the operators is negative. Thus, these Pauli operators do not appear in the HMF. This purely algebraic argument \cite{Suppmat} is independent of the temperature and the lattice geometry. 
	
	The \textit{XXZ} Hamiltonian with magnetic fields contains single-body Pauli operators and the above argument breaks down. The corresponding HMF overlaps with all Pauli operators for any $\beta>0$. 
	In table \ref{tab:Ham_terms}, we list the one and two-body terms that can arise in the HMF for several system Hamiltonians.
	
	
	\begin{table}
		\centering
		\begin{tabular}{|c|c|c|c|c|}
			\hline
			$H$ \quad & \quad $\sigma^{\alpha}_{j}$ \quad & \quad $\sigma^{\alpha}_{j}\sigma^{\alpha}_{j+1}$ \quad & \quad $\sigma^{\alpha}_{j}\sigma^{\alpha}_{k}$ \quad & \quad $\sigma^{\alpha}_{j}\sigma^{\alpha'}_{k}$ \quad \\ 
			\hline
			\textit{XX} ($\Delta=0$) \quad & \quad . \quad & \quad \checkmark \quad & \quad . \quad & \quad . \quad  \\
			\hline
			\textit{XXZ} ($\Delta\neq0$) \quad & \quad . \quad & \quad \checkmark \quad & \quad \checkmark \quad & \quad . \quad  \\
			\hline
			\textit{XXZ} $+\sigma^\alpha_n$ \quad & \quad \checkmark \quad & \quad \checkmark \quad & \quad \checkmark \quad & \quad \checkmark \quad  \\
			\hline
			\hline
			\hline
			Min.\ order in $\beta$ \quad & \quad $\beta^{2d+1}$ \quad & \quad $\beta^{2d}$ \quad & \quad $\beta^{2d}$ \quad & \quad $\beta^{2d+1}$ \quad  \\
			\hline
		\end{tabular}
		\caption{One- and two-body terms appearing in $H^{*}_{A}$, for various system Hamiltonians $H$. The final row indicates the minimum order in $\beta$ at which the terms can appear. 
	}
	\label{tab:Ham_terms}
\end{table}


\section{Small $\beta$}
For $\beta\to0$, the HMF converges to $H_A$ \cite{CresserAnders_HMFcoupling_PRL2021, ChenCheng_HMF_JocP2022}. For small $\beta>0$, we find that $H_A^*$ differs from $H_A$ most notably close to the boundary of $A$ with the bath $B$, i.e., $|c(O)|$ is larger for operators $O$ supported close to the boundary than for $O$ supported far away from the boundary.
We define the distance $d(O)$ as the distance from the boundary to the farthest site within the support of $O$, i.e., for an $n$-body Pauli operator $O=\sigma_{i_1}^{\alpha_1}\dots\sigma_{i_n}^{\alpha_n}$ with $i_1<\dots<i_n$, we have $d(O) = L_A - i_1$.

\begin{figure}
	\centering  \includegraphics[width=\linewidth]{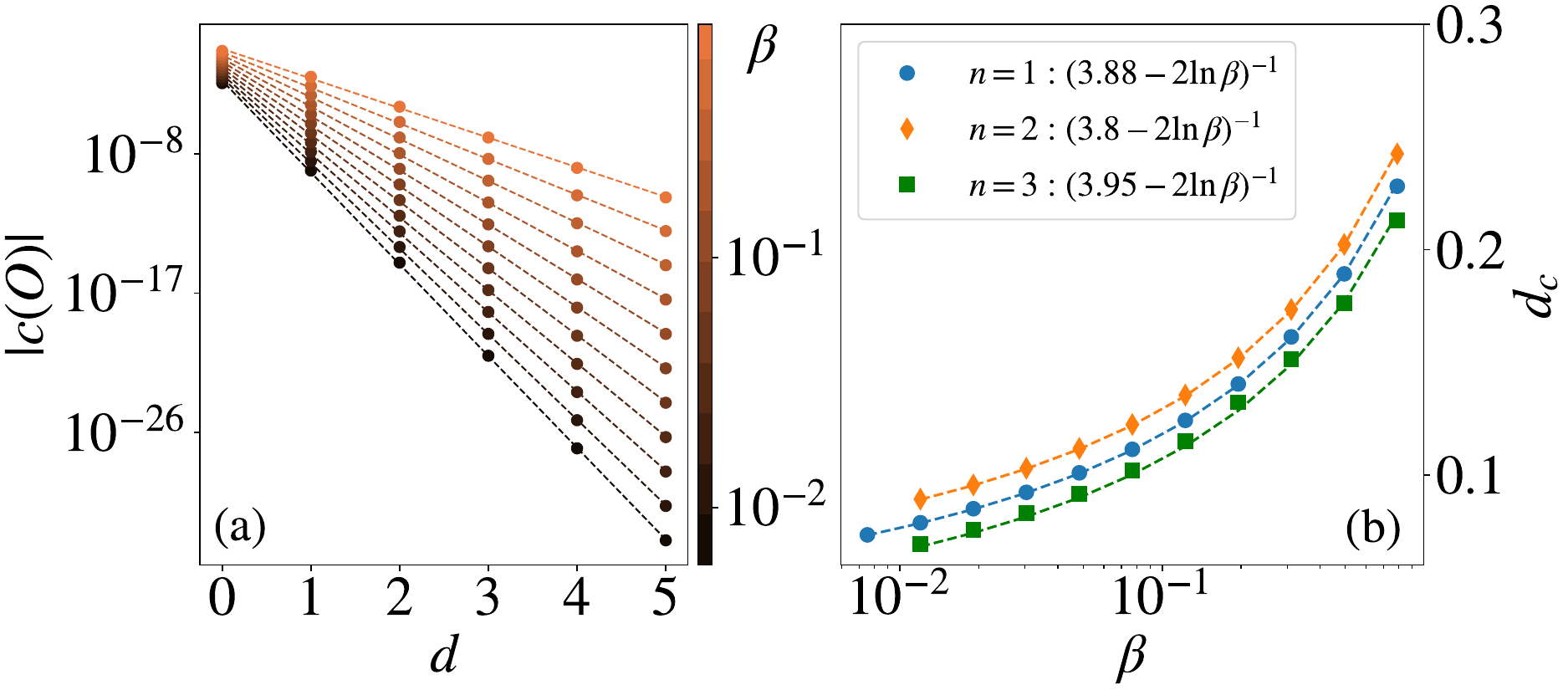}
	\caption{\label{fig:coeffs_dc_vs_beta}
		\textbf{(a)} $|c(O)|$ for 1-body term ($\sigma_j^x$) in the HMF versus distance for various $\beta$ (colorbar).  Corresponding curves for 2- or 3-body terms are similar \cite{Suppmat}. 
		\textbf{(b)} Skin depth $d_c$ as function of $\beta$, for 1-body, 2-body ($\sigma_\ell^\alpha \sigma_{m}^\alpha$), 3-body ($\sigma_\ell^\alpha \sigma_{m}^\alpha \sigma_{n}^{\gamma}$) terms. $n=1$ corresponds to slopes in (a). Dashed lines are $(a-2\ln\beta)^{-1}$ for fitted $a$ (legend) for each $n$ -- shifted up/down by $10^{-2}$ for clarity.
		Data for a uniform field \textit{XXZ} chain with $L=7$, $L_A=6$, $J=1$, $\Delta=0.95$, and $h_x = h_z = 0.2$.
	}
\end{figure}

Remarkably, $|c(O)|$ decays exponentially with distance $d$ from the boundary. 
For one-body ($n=1$) terms, the clear exponential behavior is seen in Fig.~\ref{fig:1}(b) and Fig.~\ref{fig:coeffs_dc_vs_beta} (a); similar exponential behaviors are observed for $n=2$ and $n=3$ operators \cite{Suppmat}. 
This exponential decay with distance means that the HMF differs from the bare subsystem Hamiltonian $H_A$ only by a skin effect. The skin depth $d_c$ is defined by
\begin{equation}\label{eq:dc_def}
	|c(O)| \sim e^{-d/d_c}.
\end{equation}
The skin depth is the slope in Fig.\ref{fig:1}(b) and Fig.~\ref{fig:coeffs_dc_vs_beta}(a). The skin depth as a function of $\beta$ is shown for one-, two-, and three-body terms in Fig.~\ref{fig:coeffs_dc_vs_beta}(b).  This dependence is very well approximated by
\begin{equation}\label{eq:d_c}
	d_c \approx \frac{1}{a - 2 \ln \beta}.
\end{equation}
The quantity $a(O)$ is numerically found to be either independent of or at most weakly dependent on $\beta$ and to depend on the operator $O$. The term $2\ln \beta$ is independent of the operator; we will provide a perturbative argument below.




\subsection{Perturbative considerations}
Eqs.~\eqref{eq:dc_def} and~\eqref{eq:d_c} can be justified by a perturbative argument in $\beta$. We expand the HMF in a formal power series in $\beta$,
\begin{align}\label{eq:hmf_expansion}
	H_{A}^{*} = \sum_{k=0}^{\infty} \beta^k H_{A,k}^*,
\end{align}
with the matrix-valued coefficients  \cite{Suppmat}
\begin{multline}\label{eq:hmf_expansion_H_k}
	H_{A,k-1}^* = (-1)^{k-1}
	\sum_{m=1}^{k} \frac{(-1)^{m+1}}{m\ D_B^m} \ \times 
	\\ 
	\sum_{\left\{n_1+...+n_m=k\right\}} \left(\prod_{i=1}^{m} \frac{\tr_B(H^{n_i}) }{{n_i}!}\right) \quad
	- [H\leftrightarrow H_B],
\end{multline}
where the second term, $[H\leftrightarrow H_B]$, is obtained from the first term by replacing $H$ with $H_B$. 

We note that an expansion in a `unitful' quantity $\beta$ is questionable, and indeed the expansion in Eq.~\eqref{eq:hmf_expansion} should be in $\beta/J$ as the dominating scale in the Hamiltonian is $J$, provided $h_j$ and $\Delta$ are not significantly larger than $J$. 
In all numerical results, we have chosen $J=1$, and thus for ease of readability we have opted
to omit factors of $J$.

\begin{figure}
	\includegraphics[width=\linewidth]{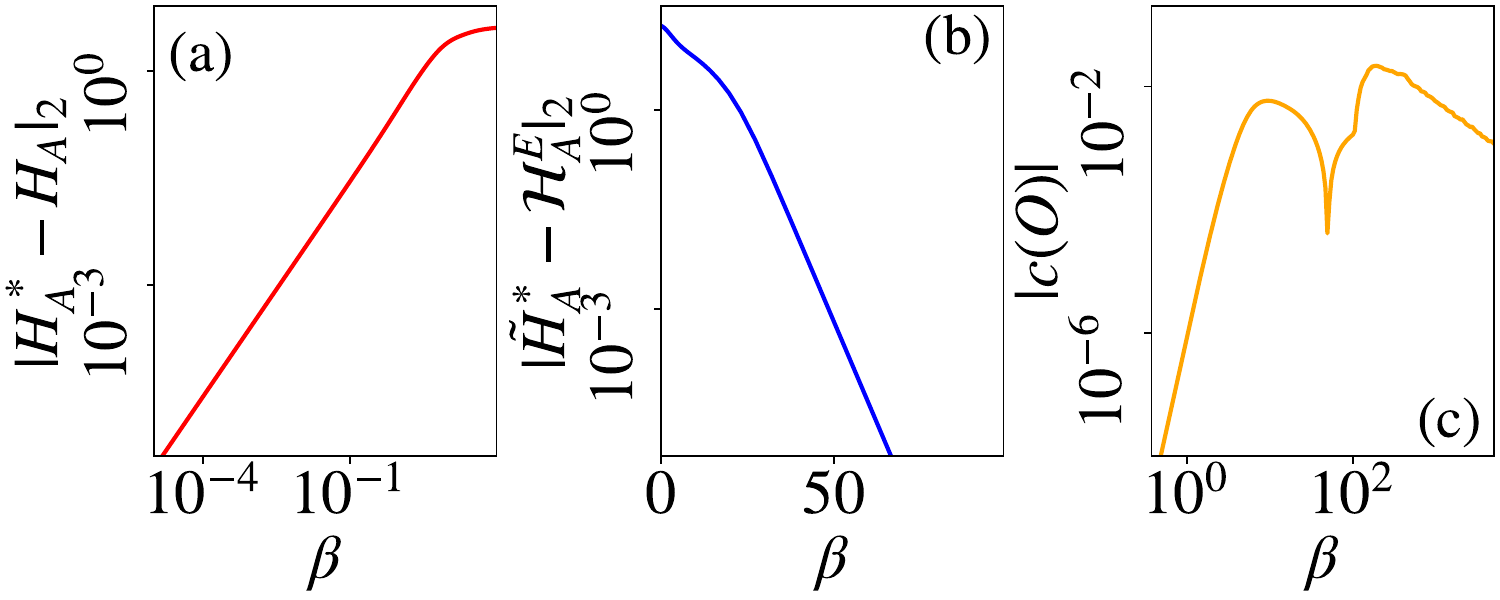}
	\caption{Hilbert-Schmidt norm of the difference between (a) $H_A$ and $H^{*}_{A}$, (b) $\mathcal{H}^{E}_A$ and $\Tilde{H}^*_A$, for a uniform field \textit{XXZ} chain with $L=7$, $L_A = 3$, $J=1$, $\Delta = 0.95$, and $h^x = h_z = 0.2$. 
		(c) $|c(O)|$ for a two-body term (that does not appear in $H_A$) in $H^{*}_{A}$ versus $\beta$, for the same parameters as (a) and (b) but with $L=6$ and $L_A=4$.}
	\label{fig_diff_HentHmf_XXZ_unih}
\end{figure}

When the operators in $H$ are trace-less, the first few coefficients are \cite{Suppmat}
\begin{align}
	H_{A,0}^* &= H_A  \label{eq:expansion_0} \\
	H_{A,1}^* &= \frac{1}{2D_B}\tr_B(H_{AB}^2)  
	\label{eq:expansion_1} \\
	H_{A,2}^* &= \frac{1}{6D_B} \left[\tr_B(H_{AB} H_A H_{AB}) - \tr_B(H_{AB}^2) H_A \right]. \label{eq:expansion_2}
\end{align}
From Eq.~\eqref{eq:expansion_0} one infers the known result $H_A^*(\beta)\to H_A$ for $\beta\to 0$ \cite{CresserAnders_HMFcoupling_PRL2021, ChenCheng_HMF_JocP2022}, which is illustrated in Fig.~\ref{fig_diff_HentHmf_XXZ_unih}(a,c). 

We denote the overlap of operators $O$ with $H_{A,k}^*$ by $c_k(O) = \tr\small(O H_{A,k}^{*}\small)$,
and the smallest $k\ge 1$ such that $c_k(O)$ is non-zero by $k_0$. 
We observed that $k_0$ is lower bounded by
\begin{equation}\label{eq:k_0_bound}
	k_0 \ge 2(d+1) - n
\end{equation} 
for $n$-body operators. In other words, the minimum order at which $c(O)$ can appear is $\mathcal{O}(\beta^{2(d+1)-n})$.
Some examples are listed in Table \ref{tab:Ham_terms}.

\begin{figure}
	\centering
	\includegraphics[width=\linewidth]{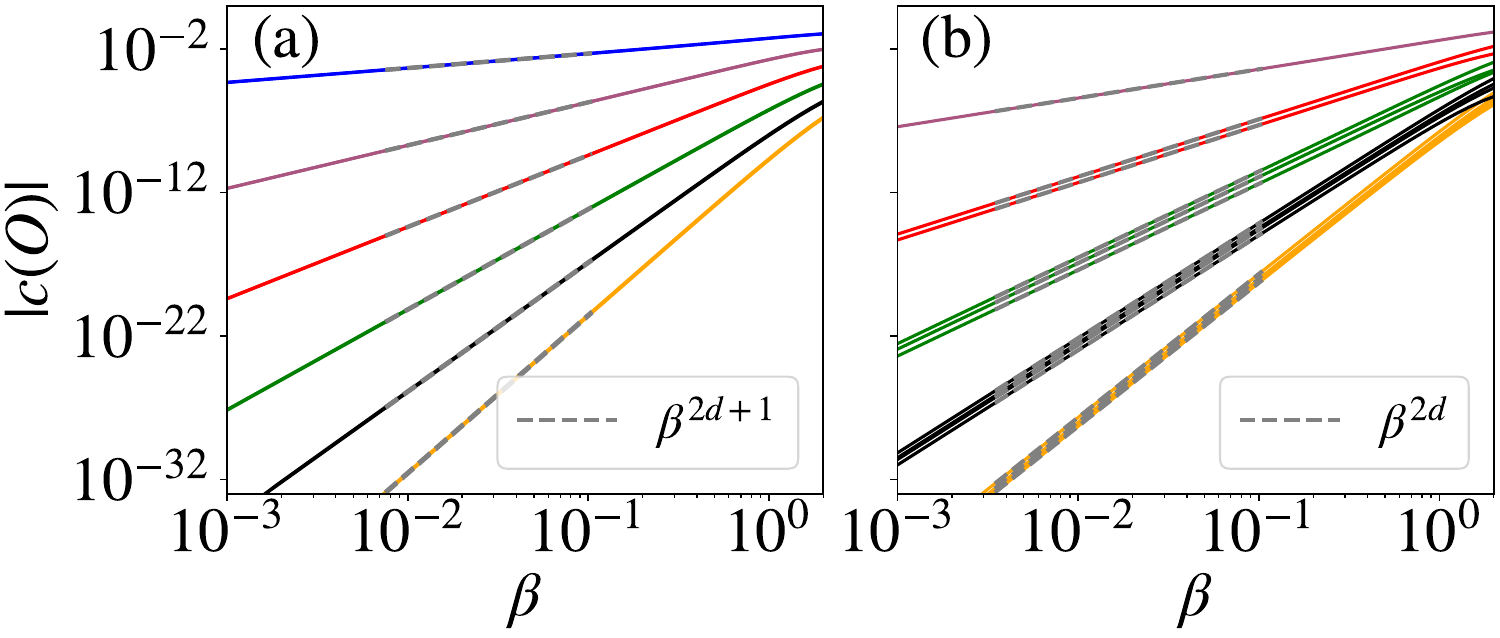}
	\includegraphics[width=\linewidth]{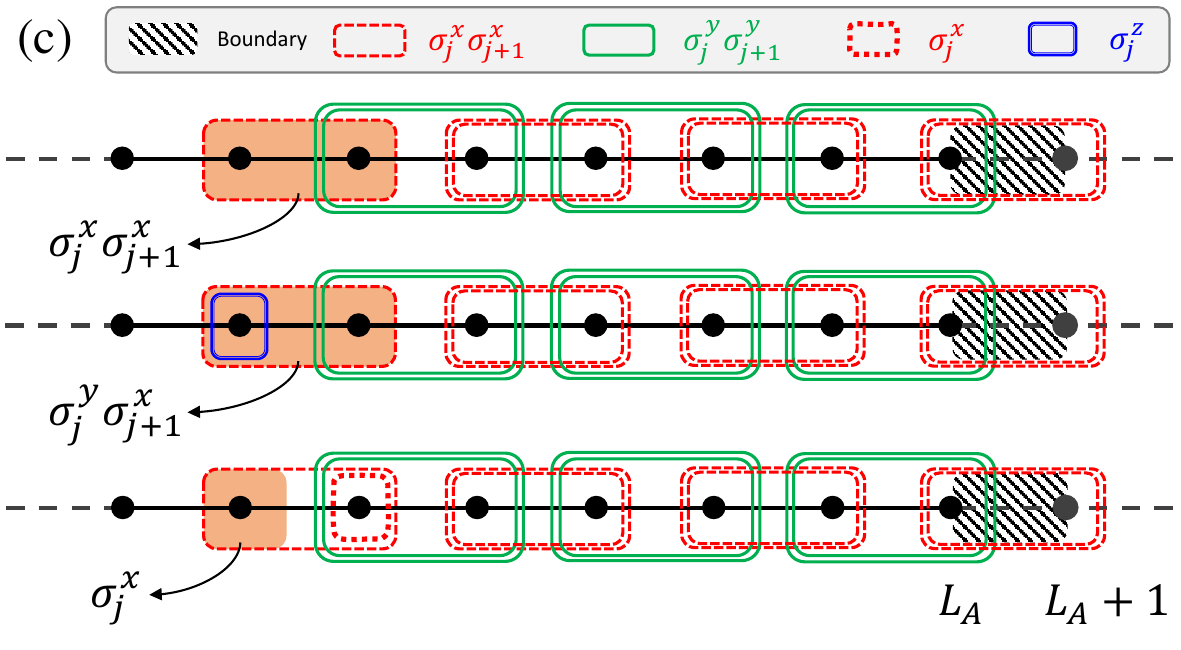}
	\caption{
		$|c(O)|$ for (a) 1-body ($\sigma_j^{(x,z)}$) and (b) 2-body ($\sigma_j^x \sigma_k^x$) operators in $H^{*}_{A}$ versus $\beta$ - Uniform field \textit{XXZ} chain with $L=7$, $L_A = 6$, $J=1$, $\Delta=0.95$, $h_x = h_z = 0.2$. Similar results are observed in an \textit{XXZ} chain with disordered fields \cite{Suppmat}. 
		Lines are grouped into colors representing the distance to the boundary, the top group being closest to the boundary.
		(c) Illustration of analytical argument for the factor of $2d$. Rectangles denote Pauli operators ($h_i$) in $H$. The different colored patterns correspond to $x,y,z$, and the enclosed black dots indicate the support of the operator on the chain. Double rectangles mean that the operators appear twice. The orange-shaded regions indicate support of operators corresponding to the product of all Pauli operators in the string ($O\equiv h_1,\dots,h_{k+1}$).
	}
	\label{fig:Coeff_vs_beta}
\end{figure}

In Fig.~\ref{fig:Coeff_vs_beta}(a,b) we present $|c(O)|$ as a function of $\beta$ for $n=1$ and $n=2$ operators.  For small $\beta$, the dependence is a power-law with exponent $2d+1$ for one-body operators and $2d$ for two-body operators $O$ (dashed-line). 
Similarly, we find that $c(O)$ is of order $\mathcal{O}(\beta^{2d-1})$ for three-body and $\mathcal{O}(\beta^{2d-2})$ for four-body operators \cite{Suppmat}.  In addition to these examples obeying the equality in Eq.~\eqref{eq:k_0_bound}, we also have cases where $k_0$ is larger than the lower bound: for mixed two-body operators $\sigma_j^\alpha\sigma_{\ell}^{\alpha'}$ with $\alpha\neq\alpha'$, $c(O)$ follows a power-law in $\beta$ with exponent $2d+1$ ($2d+2$ if $\ell=L_A$) \cite{Suppmat}.


Before justifying the lower bound, Eq.~\eqref{eq:k_0_bound}, we first highlight a consequence. 
We denote $k_0 = 2d + b$, where $b$ is an integer, independent of $d$ and $\beta$. Then
\begin{equation}\label{eq:beta_exp}
	c(O) = e^{2d\ln\beta} \sum_{k=b}^\infty c_{k+2d} \beta^{k}. 
\end{equation}
We write the $d$-dependence of the second factor as  $\left| \sum_{k=b}^\infty c_{k+2d} \beta^{k} \right| \sim e^{-ad}$, where $a$ is a real number; we assume that there is no faster dependence on $d$.  (The prefactor may have polynomial dependence.)
Then the exponential dependence of $|c(O)|$ on $d$ is given by
\begin{equation}\label{eq:pert_dc}
	|c(O)| \sim e^{-(a-2\ln\beta) d }.
\end{equation}
Extracting $d_c$ from Eq.~\eqref{eq:pert_dc} implies Eq.~\eqref{eq:d_c}.
For small $\beta$, retaining only the first term in the sum in Eq.~\eqref{eq:beta_exp}, we see that $a$ is independent of $\beta$ for small $\beta$ \cite{Suppmat}. 
Numerically (Fig.~\ref{fig:coeffs_dc_vs_beta}(b)), $a$ appears to be $\beta$-independent at all $\beta$.  



\subsection{Justification of the lower bound \eqref{eq:k_0_bound}}  

Eq.~\eqref{eq:k_0_bound} is consistent with the lowest-order expressions,  Eqs.~\eqref{eq:expansion_0}-\eqref{eq:expansion_2}. 
By Eq.~\eqref{eq:k_0_bound}, two-body Pauli operators with distance $d>1$ and one-body operators with $d\ge 1$ should have $c_k(O)=0$ for $k=1,2$. These operators are not the identity. 
But $H_{A,1}^*\propto \mathbb{1}_A$ by Eq.~\eqref{eq:expansion_1}, when $H_{AB}$ contains no mixed Pauli operators, so $c_1(O)=0$. In \cite{Suppmat} we show 
how $c_2(O)=0$ follows from Eq.~\eqref{eq:expansion_2}.
For $k>2$, $H_{A,k}^*$ becomes intractable to calculate.  However, we can formulate a heuristic argument for Eq.~\eqref{eq:k_0_bound}, based on the following


\begin{conjecture} 
	We express the operator $O$ in terms of tuples $h_1,\dots,h_{k+1}$ up to a constant, $O \equiv h_1\cdots h_{k+1}$ ($k\ge1$), where the $h_i$ are all in $H$.  If all such tuples $h_1,\dots,h_{k+1}$ can be split into two sets of operators $\mathcal{H}_1\neq\emptyset$ and $\mathcal{H}_2$, such that operators in $\mathcal{H}_1$ commute with operators in $\mathcal{H}_2$ and operators in $\mathcal{H}_1$ have no support in $B$, then $c_{k}(O)=0$. 
\end{conjecture}

The conjecture is supported by the $k=1$ and $k=2$ cases, which follow from Eqs.\ \eqref{eq:expansion_1} and \eqref{eq:expansion_2} \cite{Suppmat}.  



As shown visually in Fig.~\ref{fig:Coeff_vs_beta}(c), the smallest $k$ such that there exists a string $h_1\dots h_{k+1}\equiv O$, which cannot be separated into $\mathcal{H}_{1,2}$ is given by $k=2d$ for nearest-neighbor two-body operators $O=\sigma_j^\alpha\sigma_{j+1}^\alpha$ and $k=2d+1$ for single-body operators $O$ and mixed nearest-neighbor two-body operators $O=\sigma_j^\alpha\sigma_{j+1}^{\alpha'}$ with $\alpha\neq\alpha'$. These strings are given by nearest neighbor two-body operators appearing in $H$ connecting the bath to the support of the operator $O$. 
We sketch such strings in Fig.~\ref{fig:Coeff_vs_beta}(c) for $O=\sigma_j^x\sigma_{j+1}^x$, $O=\sigma_j^y\sigma_{j+1}^x$  and $O=\sigma_j^x$. 

The first row of Fig.~\ref{fig:Coeff_vs_beta}(c) corresponds to representations of $O=\sigma_j^x\sigma_{j+1}^x$ of the form
\begin{equation}\label{eq:string}
	\sigma_j^x\sigma_{j+1}^x \equiv \sigma_{j}^x\sigma_{j+1}^x \big(\sigma_{j+1}^y \sigma_{j+2}^y\big)^2 \dots \big(\sigma_{L_A}^{(x/y)} \sigma_{L_A+1}^{(x/y)}\big)^2
\end{equation}
or permutations thereof. 
Using squares of pairs requires the least number of Pauli operators that connect to the bath while yielding the identity operator. 
The square structure in Eq.\ \eqref{eq:string}, reflected in the double rectangles in Fig.~\ref{fig:Coeff_vs_beta}(c), provides a visual interpretation of the factor 2 in the lower bound \eqref{eq:k_0_bound}. 

The strings for $O=\sigma_j^y\sigma_j^x$ and $O=\sigma_j^x$, also presented in Fig.~\ref{fig:Coeff_vs_beta}(c) are similar. They differ only by the first term on the right-hand side in Eq.~\eqref{eq:string}, namely $\sigma_j^y\sigma_{j+1}^x\equiv \sigma_j^z \sigma_j^x\sigma_{j+1}^x$ and $\sigma_j^x \equiv \sigma_{j+1}^x\sigma_j^x\sigma_{j+1}^x$. This agrees with our result that single-body and mixed two-body terms only appear in the HMF when magnetic fields are present in $H$ \cite{Suppmat}.

For nearest neighbor two-body operators, $\sigma_j^\alpha \sigma_{j+1}^\alpha$, any Pauli string $h_1\dots h_{k+1}$ representing the operator with $k<2d$ can be separated into sets $\mathcal{H}_{1,2}$, because there are not enough operators to construct structures of the type shown in Fig.~\ref{fig:Coeff_vs_beta}(c). Similarly, for $k<2d+1$ this holds for single-body operators and nearest neighbor two-body operators $\sigma_j^\alpha \sigma_{j+1}^{\alpha'}$ with $\alpha\neq\alpha'$.

\section{Large $\beta$}

In the limit $\beta\to\infty$, $e^{-\beta H} /Z$ converges to the projector $P_{GS}=\ket{\Psi_{GS}}\bra{\Psi_{GS}}$ onto the ground state subspace, where $\ket{\Psi_{GS}}$ is the ground state wavefunction of the entire ($A+B$) system. The reduced projector, 
\begin{equation}
	\rho_A^{GS}=\tr_B(P_{GS}) = \tr_B \outerprod{\Psi_{GS}}{\Psi_{GS}}, 
\end{equation}
i.e., the reduced density matrix of the $A$ region in the ground state wavefunction, is a widely studied object, as it encodes the entanglement between $A$ and $B$.  It is often expressed in terms of an `effective' Hamiltonian $\mathcal{H}^{E}_A$, the entanglement Hamiltonian \cite{LiHaldane_EntanglementSpectrum_PRL2008, PeschelEisler_RDMEnt_JoPA2009, EislerPeschel_PropEntHam_JoSM2018, DalmonteEisler_EntHamRev_AndP2022}:
\begin{equation}
	e^{-\mathcal{H}^{E}_A} = \rho_A^{GS} = \lim_{\beta\to\infty} \tr_B (e^{-\beta H}) Z^{-1}. 
\end{equation}
Comparing with Eq.\ \eqref{eq:hmf_def}, we obtain the following relationship for  $\beta \gg 1$ between the HMF $H_A^{*}$ and the entanglement Hamiltonian $\mathcal{H}^{E}_A$:  
\begin{equation}
	\mathcal{H}^{E}_A \approx \beta H_A^{*} + \ln(Z^*).
\end{equation}
In other words, the entanglement Hamiltonian is obtained by shifting and scaling the HMF at large $\beta$.  
This relation extends the previously derived result that $H_{A}^{*}\propto \mathbb{1}_A$ in the same limit \cite{ChenCheng_HMF_JocP2022}.

The shifted and scaled operator  $\Tilde{H}^*_A =\beta H_A^{*} + \ln(Z^*)$ is compared in Fig.~\ref{fig_diff_HentHmf_XXZ_unih}(b) to $\mathcal{H}^{E}_A$ for $\beta \gg 1$ limit.  As predicted, the distance between the two `Hamiltonians'  vanishes for $\beta\to\infty$.

\section{Concluding discussion}
In this work, we investigated the (spatial) structure of the HMF, a fundamental concept in understanding the implications of non-weak coupling in thermodynamics. 
We demonstrated and explained a skin effect in the HMF structure: The magnitude of terms in $H_A^*-H_A$ decreases exponentially with the distance from the boundary. 
We also identified the types of terms that can appear in the HMF and at which order they can occur. 

The idea of a skin effect in the deviation of the HMF from the system Hamiltonian is related in spirit to ideas discussed in the literature around locality of temperature \cite{KlieschEisert_LocalityOfTemp_PRX2014, Hovhannisyan_LocalityOfTemp_NJoP2015, DePalma_UniversalLocality_PRA2017, Alhambra_QMB_ThermalEquilibrium_PRXQ2023}.  
Since thermal states of systems described by short-range interactions incorporate a notion of locality, it makes sense that the effect on the HMF should be localized near the boundary. 
To the best of our knowledge, one cannot infer the explicit exponential behavior, Eqs.\ \eqref{eq:dc_def} and \eqref{eq:d_c}, from such intuition alone.  

While our explicit examples and expressions are for a spin-$\frac{1}{2}$ chain with one-body and nearest-neighbor two-body terms and for traceless operators, it is clear that analogous expressions can be worked out for other cases (fermions, bosons, spins $>\frac{1}{2}$, other local operators, other geometries), and that the physical conclusions are generic. 

The form of the skin effect and the skin depth, Eqs.~\eqref{eq:dc_def} and \eqref{eq:d_c}, do not depend on the strength of the ``system-bath'' coupling $H_{AB}$. 
We see from the perturbative construction that the coupling strength can affect the coefficient at most polynomially in $d$, leaving the exponential decay in Eq.~\eqref{eq:d_c}  and hence the length $d_c$ unaffected.  We have also explicitly tested this independence numerically \cite{Suppmat}. 

Our perturbative construction, and the arguments leading to the skin effect, do not require the subsystem $B$ to be large.  Thus, the results are independent of `bath size' in the sense that the form $|c(O)| \sim e^{-d/d_c}$ of  Eq.~\eqref{eq:dc_def}, and the value of 
$d_c$, are not affected by the size $L_B$ of the $B$ partition \cite{Suppmat}.  Changing $L_B$ does affect the prefactor in  Eq.~\eqref{eq:dc_def}, i.e., the magnitude of $|c(O)|$.  

In addition, the chaotic vs. integrable nature of the Hamiltonian, although generally important for thermalization/thermodynamics, plays no role, and our results are independent of integrability. 

Our results open up several research directions.  
(1) The skin effect structure is based on locality; one might ask whether some version of this picture survives for long-ranged Hamiltonians with power-law decay of couplings. 
(2) Notions of boundary locality have been discussed for the entanglement Hamiltonian and its spectrum \cite{AlbaHaqueLauchli_BoundaryLocality_PRL2012, AlbaHaqueLauchli_BoseHubbard_PRL2013, DalmonteEisler_EntHamRev_AndP2022}.  It is intriguing to ask whether these might be related to the skin effect we have presented for the HMF.
(3) Numerically, we found the constant $a$ in Eq.~\eqref{eq:d_c} to be relatively large ($>1$), leading to a rather sharp skin effect (small $d_c$); the deviations of $H_A^*$ from $H_A$ are strongly localized near the boundary. Whether this is a generic feature for different classes of systems remains an open question. 
(4) In this work we have focused on the exponential behavior $|c(O)| \sim e^{-d/d_c}$ and on the constant $d_c$, and have not attempted to treat the prefactor, i.e., the absolute magnitude of $|c(O)|$, explicitly.  For example, the way this prefactor depends on the size of the $B$ partition remains an open question.

\section*{Acknowledgments}
PCB acknowledges funding from Science Foundation Ireland through grant 21/RP-2TF/10019. This work was in part supported by the Deutsche Forschungsgemeinschaft under grant SFB 1143 (project-id 247310070). 

\normalem
%



\setcounter{page}{1} \renewcommand{\thepage}{S\arabic{page}}

\setcounter{figure}{0}
\renewcommand{\thefigure}{S\arabic{figure}}

\setcounter{equation}{0}
\renewcommand{\theequation}{S.\arabic{equation}}

\setcounter{table}{0}
\renewcommand{\thetable}{S.\arabic{table}}

\setcounter{section}{0}
\renewcommand{\thesection}{S.\Roman{section}}

\renewcommand{\thesubsection}{S.\Roman{section}.\Alph{subsection}}

\makeatletter
\renewcommand*{\p@subsection}{}
\makeatother

\renewcommand{\thesubsubsection}{S.\Roman{section}.\Alph{subsection}-\arabic{subsubsection}}

\makeatletter
\renewcommand*{\p@subsubsection}{}  
\makeatother

\vspace*{0.25cm}
\begin{center}
	\large{
		Supplemental Material for \\ \textit{Structure of the Hamiltonian of mean force}
	}
\end{center}

	\section{Overview}
	
	In the Supplemental Material, we provide supporting information and data. 
	\begin{itemize}
		\item In Section \ref{supp_zero_coeff} we explain which terms can and cannot appear in the Hamiltonian of mean force for the \textit{XXZ} chain.
		\item In Section \ref{supp_perturb} we provide the small $\beta$ expansion of the HMF. We further prove the ``lower bound'' and the conjecture in the main text for $k=1$ and $k=2$.
		\item In Section \ref{supp_nbody} we provide additional data for different $n$-body coefficients in the HMF.
		\item In Section \ref{supp_disorder} we present some data for a disordered field model.
		\item In Section \ref{supp_tuningJab} we show the result of tuning the coupling $J_{AB}$.
	\end{itemize}
	

	
	\section{Zero coefficients of \textit{XXZ} model without magnetic fields}\label{supp_zero_coeff}
	In this section, we show that single-body and two-body mixed Pauli operators do not appear in the HMF of the \textit{XXZ} model without magnetic fields. The result is independent of the geometry of the underlying lattice. 
	
	Let us first define a `sign' of a Pauli operator $\sigma^\alpha$, where $\alpha\in\{0,x,y,z\}$,
	\begin{equation}
		s(\sigma^{\alpha}) = \begin{cases}
			1  & \alpha=0 \text{ or } x\\
			-1 & \alpha=y \text{ or } z.
		\end{cases}
	\end{equation}
	The choice of which operator to assign $+1$ is arbitrary. The identity has to have $s=+1$. The two other Pauli operators will be assigned $s=-1$. In the following we will interpret the spin algebra $\mathfrak{su}(2)$ with the Klein four-group by identifying $\sigma^x\sigma^y \equiv \sigma^y \sigma^x \equiv \sigma^z$ and equivalently for the other Pauli operators. Then it is easy to see that $s$ is a homomorphism on the Klein group, i.e. for $\alpha, \alpha'\in\{0,x,z,y\}$,
	\begin{equation}
		s(\sigma^{\alpha} \sigma^{\alpha'} ) = s(\sigma^\alpha) s(\sigma^{\alpha'}).
	\end{equation}
	This homomorphism is generalized to Pauli strings via
	\begin{equation}
		s\left( \prod_i \sigma_i^{\alpha_i} \right) = \prod_i s(\sigma_i^{\alpha_i}).
	\end{equation}
	
	Obviously, all Pauli operators in the \textit{XXZ} Hamiltonian $H_{XXZ}$ without magnetic fields have $s=+1$. By the homomorphism property of $s$, this generalizes to all Pauli operators in powers of $H_{XXZ}$. Together with the fact that $\exp$ is an analytic function, this implies that $\exp(-\beta H_{XXZ})$ only contains Pauli operators with $s=+1$.
	
	The partial trace $\tr_B$ is a projection, such that for every Pauli operator $\boldsymbol{\sigma}$, $\tr_B(\boldsymbol{\sigma}) = \boldsymbol{\sigma}$, if and only if $\boldsymbol{\sigma}$ has no support on $B$, and $\tr_B(\boldsymbol{\sigma})=0$ otherwise. Especially, $\tr_B$ preserves $s$, whenever the Pauli operator $\boldsymbol{\sigma}$ is not projected onto 0. Thus, $\tr_B(\exp(-\beta H_{XXZ}))$ only contains Pauli operators with $s=+1$.
	
	Again, by exploiting the homomorphism property of $s$ and that $\log$ is an analytic function, all Pauli operators in the HMF of the \textit{XXZ} model without magnetic fields have $s=+1$. This excludes especially $\sigma^{y,z}$ and $\sigma_i^x\sigma_j^y$, $\sigma_i^x\sigma_j^z$ for $i\neq j$. 
	
	The above arguments are invariant under interchanging $x,y,z$ arbitrarily while keeping the assignment of signs such that one Pauli operator has $s=+1$ and the two other non-trivial ones have $s=-1$. Thus the HMF of the \textit{XXZ} model without magnetic fields does not contain any single-body terms nor mixed two-body terms $\sigma_i^\alpha\sigma_j^{\alpha'}$ with $\alpha\neq\alpha'$. Further, 3-body and 4-body terms with positive signs for any sign function like $\sigma^x\sigma^y\sigma^z$ and $\sigma^x\sigma^y\sigma^x\sigma^y$ do appear, while terms like $\sigma^x\sigma^x\sigma^y$ or $\sigma^x\sigma^y\sigma^x\sigma^z$, which can have negative signs, do not. 
	
	
	\section{Small $\beta$ expansion of the HMF}\label{supp_perturb}
	
	
	In this section, we derive the expansion of the HMF around $\beta=0$ and the explicit forms of the matrix-valued coefficients $H_{A,k}^*$ for $k=0,1,2$, utilized in the main text.
	
	Recall the definition of the HMF, presented in the main text,
	\begin{equation}
		H^{*}_{A}(\beta) = -\frac{1}{\beta} \ln \frac{\tr_B(e^{-\beta H})}{\tr_B(e^{-\beta H_B})}.
	\end{equation}
	We can separate the logarithm into two terms, as the denominator is simply a number, giving
	\begin{equation}\label{supp_eq_hmf_begin}
		H_{A}^{*} = -\beta^{-1} \left( \ln(\tr_{B}e^{-\beta H})- \ln(\tr_{B}e^{-\beta H_B})\cdot\mathbb{1}_A\right).
	\end{equation}
	Now, to evaluate this, we shall write down the full power series of the matrix exponential 
	\begin{align}
		\tr_B e^{-\beta H} \ &= \ \tr_B \sum_{n=0}^{\infty} \frac{(-\beta H)^n}{n!} \notag \\
		&= \ D_B \mathbb{1}_A + \sum_{n=1}^{\infty} \tr_B\frac{(-\beta H)^n}{n!}
	\end{align}
	and logarithm
	\begin{align}
		\ln(\mathbb{1} + X)\ &=\ \sum_{n=1}^{\infty} \frac{(X)^m (-1)^{m+1}}{m}.
	\end{align}
	Following some algebra, one obtains for the first term in Eq.~\eqref{supp_eq_hmf_begin}
	\begin{align}
		\ln(\tr_{B}e^{-\beta H}) 
		=&\sum_{k=1}^{\infty} (-\beta)^{k} \sum_{m=1}^{k} D_B^{-m}\frac{(-1)^{m+1}}{m} \notag\\
		&\sum_{\left\{n_1+...+n_m=k\right\}} \left(\prod_{i=1}^{m} \frac{\tr_B(H^{n_i}) }{{n_i}!}\right),
	\end{align}
	where the last sum is over all integers $n_1,\dots,n_m \ge 1$ with $n_1+\dots+n_m=k$. The second term is retrieved similarly by replacing $H$ with $H_B$. 
	Thus, the Hamiltonian of mean force is
	\begin{align}
		H_{A}^{*} \ = \ \sum_{k=1}^{\infty} &(-\beta)^{k-1} \sum_{m=1}^{k} D_B^{-m}\frac{(-1)^{m+1}}{m} \notag\\
		&\sum_{\left\{n_1+...+n_m=k\right\}} \left(\prod_{i=1}^{m} \frac{\tr_B(H^{n_i}) }{{n_i}!}\right) \notag\\
		- \ \sum_{k=1}^{\infty} &(-\beta)^{k-1} \sum_{m=1}^{k} D_B^{-m}\frac{(-1)^{m+1}}{m} \notag\\
		&\sum_{\left\{n_1+...+n_m=k\right\}} \left(\prod_{i=1}^{m} \frac{\tr_B(H_{B}^{n_i}) }{{n_i}!}\right) \notag \\
		&=\sum_{k=0}^\infty \beta^k H_{A,k}^*,
	\end{align}
	with
	\begin{align}
		H_{A,k-1}^{*} \ 
		& = \ (-1)^{k-1}
		\sum_{m=1}^{k} \frac{(-1)^{m+1}}{m\ D_B^m} \cdot \notag\\
		\sum_{\left\{n_1+...+n_m=k\right\}} & \Bigg[ 
		\prod_{i=1}^{m} \frac{\tr_B(H^{n_i}) }{{n_i}!} - \prod_{i=1}^{m} \frac{\tr_B(H_B^{n_i}) }{{n_i}!}\Bigg]
	\end{align}
	Proceeding, we assume that the terms in the Hamiltonian $H$ are traceless. This is of course true for spin-1/2 systems.
	
	The zeroth order term is then calculated as
	\begin{align}
		H_{A,0}^* 
		&= \frac{1}{D_B} \left( \tr_B(H) - \tr_B(H_B)\right)
		= H_A,
	\end{align}
	where we used $\tr_B(H_{AB})=0$. The first-order term is given by
	\begin{align}
		H_{A,1}^* &= -\left( \frac{\tr_B(H^2)}{2 D_B} - \frac{\tr_B(H_B^2)}{2 D_B} - \frac{(\tr_BH)^2}{2 D_B^2} \right) \notag \\ 
		&= \frac{1}{2}\left( H_A^2 +\frac{tr_B(H_B^2)}{D_B} - \frac{D_B H_A^2 + \tr_B (H_B^{2}+H_{AB}^2)}{D_B} \right) \notag \\
		&= \frac{1}{2 D_B}\tr_B (H_{AB}^2).
	\end{align}
	If the boundary term $H_{AB}$ does only contain non-mixed terms $\sigma_i^\alpha \sigma_j^\alpha$ for $\alpha\in\{x,y,z\}$ then $\tr_B(H_{AB}^2) \propto \mathbb{1}_A$, so we get
	\begin{align}
		H_{A,1}^* & \propto \mathbb{1}_A.
	\end{align}
	
	The second-order term is given by
	\begin{align}
		H_{A,2}^* &= -\Bigg[
		\left( \frac{\tr_{B} H^3}{3!D_B} \right) +
		\left( \frac{(\tr_BH)^3}{3D_B^3} \right) - \notag \\
		&\frac{1}{2 D_B^2}\left( \tr_BH\cdot\tr_BH^2/2 + \tr_BH^2\cdot\tr_BH/2 \right) 
		\Bigg] \notag\\ 
		&= \left[
		\frac{H_A^3}{3} + \frac{\tr_B H^3}{6D_B} - \frac{1}{4D_B}\left[ H_A, \tr_BH^2 \right]_{+}
		\right] \notag\\ 
		&= \left[ \frac{4H_A^3}{12} + \frac{2\tr_B H^3}{12D_B} - \frac{3}{12D_B}\left[ H_A, \tr_BH^2 \right]_{+}
		\right],
	\end{align}
	where $[A,B]_+ = AB + BA$ is an anticommutator. The first two terms will contain $\frac{6}{12}H_A^3$ and the anticommutator will contain $\frac{3}{12}\cdot2H_A^3$, so all powers of $H_A$ cancel. Evaluating the term we get up to a constant shift
	\begin{align}
		H_{A,2}^* &= \frac{1}{6D_B} \left( \tr_B(H_{AB} H_A H_{AB}) - \tr_B(H_{AB}^2) \cdot H_A \right).
	\end{align}

	\subsection{Lowest-order expressions and the ``lower bound''}
	In the following, we show that the lower bound on $k$ such that $c_k(O)=\tr(OH_{A,k}^*)\neq 0$ is consistent with the explicit forms of $H_{A,k}^*$ for $k=1,2$. Namely, we show for Pauli operators with $d(O) > 1$ ($d(O)\ge 1$ for single-body Pauli operators) that $c_k(O)=0$ for $k=1,2$. 
	
	First, the statement is trivial for $k=1$ as $H_{A,1}^*$ is proportional to the identity operator, whenever $H_{AB}$ does not contain mixed Pauli operators.
	
	So let us turn to $H_{A,2}^*$. We expand $H_A$ in terms of operators $h_A$, which are proportional to Pauli operators, $H_A = \sum_{h_A} h_A$. By separating the sum over $h_A$'s into $h_A$'s commuting and not commuting with $H_{AB}$ one easily sees that only the non-commuting $h_A$'s are contributing to $H_{A,2}^*$,
	\begin{equation}
		\sum_{h_A:[h_A, H_{AB}]=0}
		\left[ \tr_B(H_{AB} h_A H_{AB}) - \tr_B(H_{AB}^2) \cdot h_A \right]=0,
	\end{equation}
	so $c_2(O)$ is proportional to
	\begin{equation}\label{supp_eq_c2}
		\sum_{h_A:[h_A, H_{AB}]\neq0}
		\tr(O[ \tr_B(H_{AB} h_A H_{AB}) - \tr_B(H_{AB}^2) \cdot h_A]).
	\end{equation}
	Recall that $\tr(O_1O_2)\neq 0$ for two Pauli operators $O_1,O_2$ implies that $O_1$ equals $O_2$ up to a multiplicative factor. Especially, the supports of both operators have to coincide. 
	
	Let us now restrict to Hamiltonians $H$ with only nearest neighbor terms. Then all $h_A$ not commuting with $H_{AB}$ have support contained in $[L_A-1, L_A]$. The same holds for the supports of $\tr_B(H_{AB}h_A H_{AB})$ and $\tr_B(H_{AB}^2) \cdot h_A$. But for a Pauli operator $O$ with $d(O)>1$, the support of $O$ is by definition of the distance $d$ not contained in $[L_A-1,L_A]$. Thus the support of $O$ does not coincide with the support of $\tr_B(H_{AB}h_A H_{AB})$ nor with the support of $\tr_B(H_{AB}^2) \cdot h_A$. By the argument in the previous paragraph, $\tr(O\tr_B(H_{AB} h_A H_{AB})) = \tr(O\tr_B(H_{AB}^2) \cdot h_A)=0$, so $c_2(O)=0$.
	
	Finally, consider a single-body Pauli operator $O$ with $d(O)=1$. Then $O$ is supported only on-site $L_A-1$ and commutes with $H_{AB}$. If the boundary term $H_{AB}$ contains only non-mixed terms $\sigma_i^\alpha\sigma_j^\alpha$ for $\alpha\in\{x,y,z\}$ then $\tr_B(H_{AB}h_AH_{AB}) \propto \tr_B(H_{AB}^2)h_A \propto h_A$. But the sum in Eq.~\eqref{supp_eq_c2} only runs over $h_A$, which are \textit{not} commuting with $H_{AB}$. Thus $O$ and $h_A$ are not proportional and so $c_2(O)=0$.
	
	\subsection{Conjecture for $k=1,2$}
	In this subsection, our objective is to demonstrate that the conjecture presented in the main text holds for cases where $k=1$ and $k=2$. We restrict to operators $O$ which are supported in $A$ as otherwise $c_k(O)=0$ holds trivially for all $k$.
	
	Let us first consider $k=1$ and an operator $O$ that meets the conjecture's prerequisites. It follows that $O$ cannot be the identity operator, as the identity can always be expressed as $\operatorname{Id}=h_{AB}^2$, where $h_{AB}$ is any Pauli operator in $H$ with support in $A$ and $B$. In the absence of any mixed terms within $H_{AB}$, it follows that $H_{A,1}^* \propto \tr_B(H_{AB}^2) \propto \operatorname{Id}$, leading to the conclusion that $c_1(O)=\tr(O H_{A,1}^*)=0$.
	
	Now, let us consider $k=2$. Our goal is to demonstrate that an operator $O$ fulfilling the conjecture's conditions implies $d(O)>1$ ($d(O)\ge 1$ for single-body operators $O$). In that case, the preceding subsection's arguments imply that $c_2(O)=0$. We will verify the previous statement through contra-position: if $O$ is a single-body operator and $d(0)=0$, then $O$ is supported on $L_A$. Conversely, if $O$ is not a single-body operator and $d(O)=1$ then $O$ is a two-body operator and has support on $L_A$. Take any term $h_{AB}$ in $H$, which is supported on the boundary of $A$ and $B$ and does not commute with the single-body Pauli operator of $O$ on $L_A$. Then $O\equiv Oh_{AB}^2$ cannot be decomposed into the set $\mathcal{H}_1$ and $\mathcal{H}_2$ as outlined in the conjecture. This completes the proof.
	
	\section{Additional data for n-body terms}\label{supp_nbody}
	
	Here, we provide further data for different $n$-body terms for the `uniform field' \textit{XXZ} chain, complementing what is already presented in the main text.
	
\begin{figure}
		\centering
		\includegraphics[width=\linewidth]{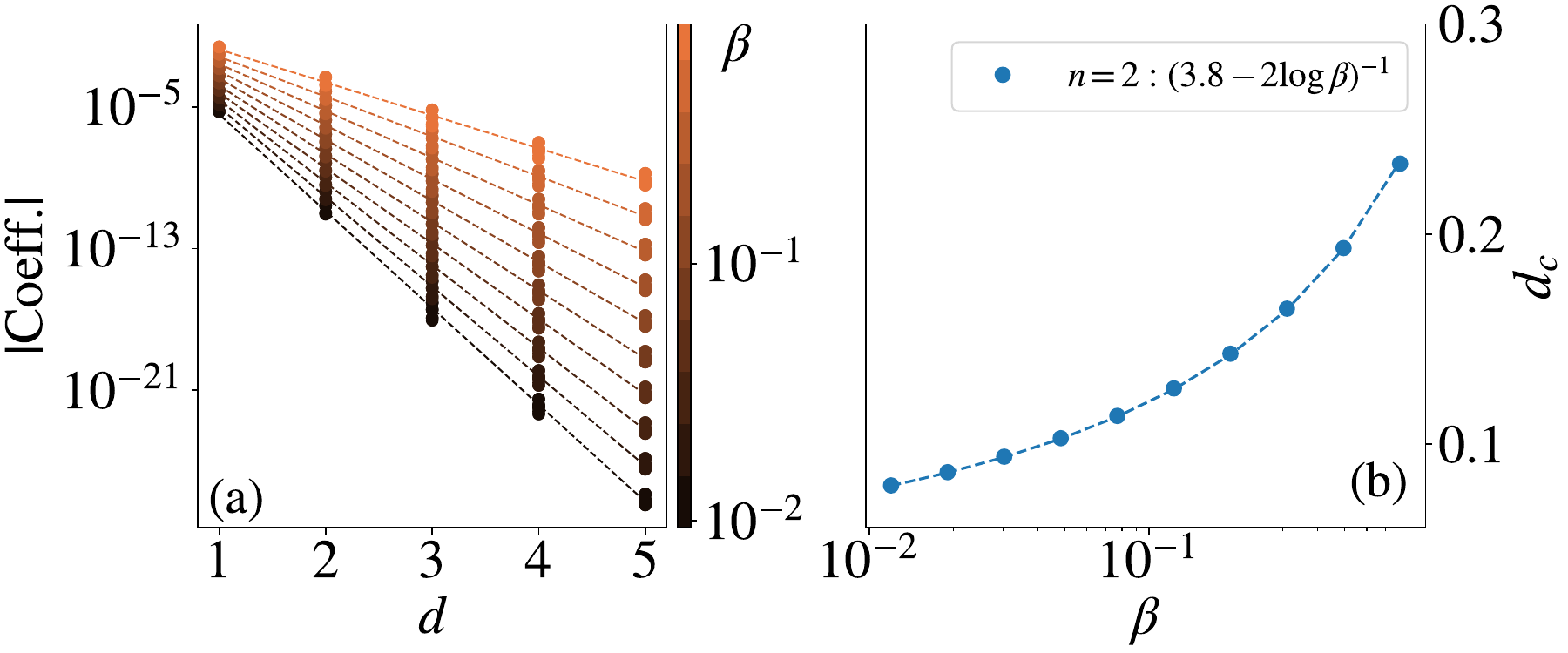}
		\includegraphics[width=\linewidth]{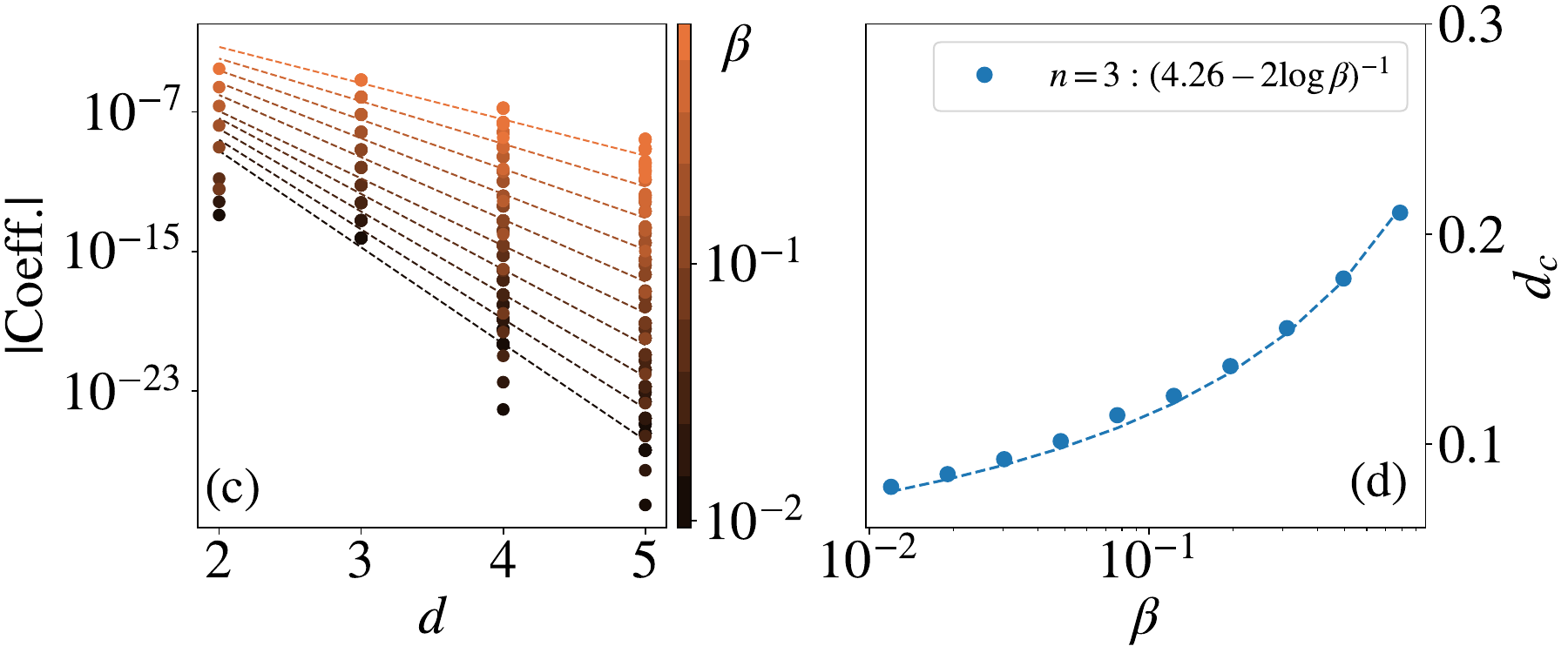}
		\caption{Magnitude of (a) 2-body ($\sigma_j^\alpha \sigma_k^\alpha$) and (c) 3-body ($\sigma_j^\alpha \sigma_k^\alpha \sigma_\ell^{\alpha'}$) coefficients in the Hamiltonian of mean force plotted versus distance over a range of $\beta$. (b,d) The slope of the lines of best fit to the data in (a,c) is plotted as a function of $\beta$. -- Uniform field \textit{XXZ} chain with $L = 7$, $L_A = 6$, $J=1$, $\Delta=0.95$, $h_x = h_z = 0.2$.}
		\label{supp_fig_coeff_vs_d_dc}
	\end{figure}

	Firstly, in Fig.~\ref{supp_fig_coeff_vs_d_dc}(a,c), we plot the coefficients of two and three-body terms as a function of the distance $d$ to the boundary. In (b,d), we plot the value of $d_c$ as a function of $\beta$, again illustrating the form $d_c = (a-2\log\beta)^{-1}$ presented in the main text.
	
\begin{figure}
		\centering
		\includegraphics[width=\linewidth]{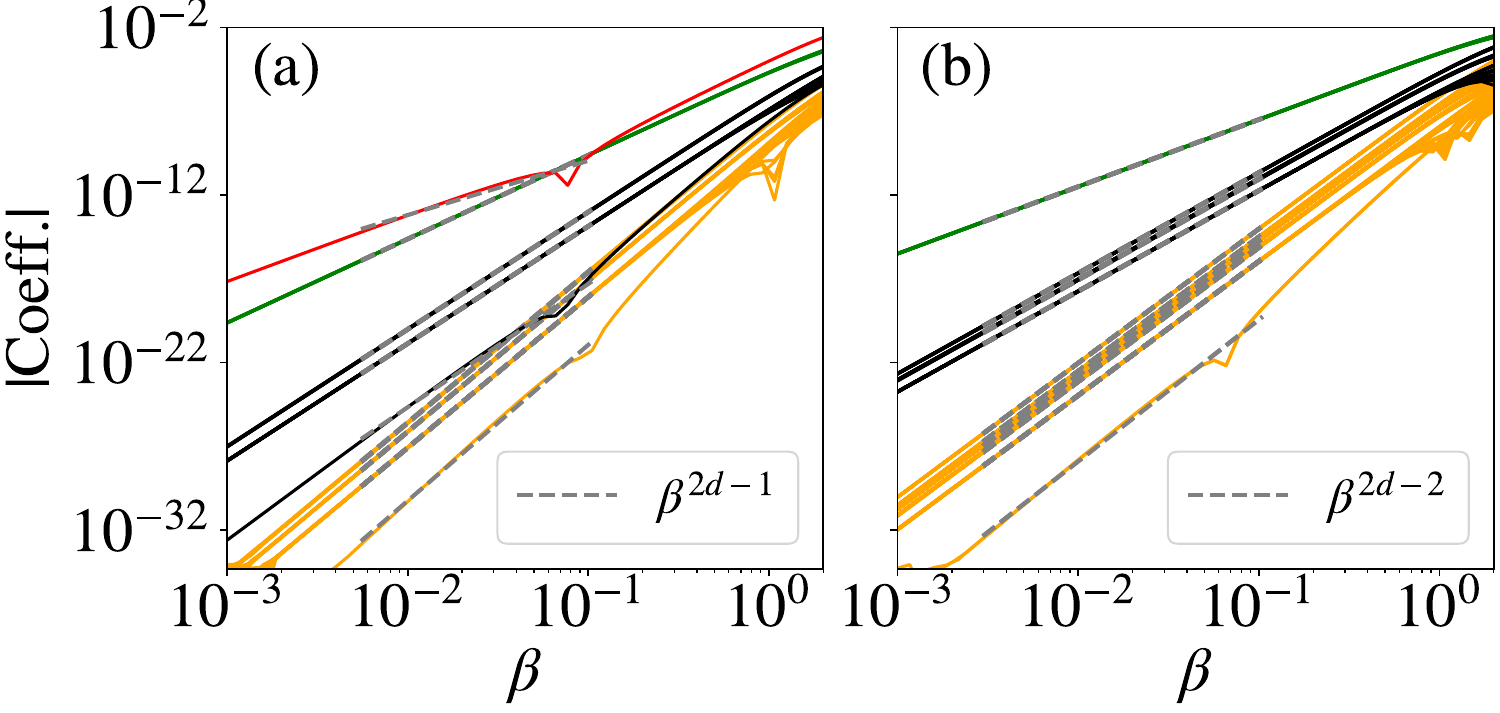}
		\caption{Magnitude of (a) 3-body ($\sigma_j^{\alpha}\sigma_k^{\alpha}\sigma_\ell^{\alpha'}$) and (b) 4-body ($\sigma_j^{\alpha}\sigma_k^{\alpha}\sigma_\ell^{\alpha'}\sigma_m^{\alpha'}$) coefficients in the Hamiltonian of mean force plotted versus $\beta$ - Uniform field \textit{XXZ} chain with $L=7$, $L_A = 6$, $J=1$, $\Delta=0.95$, $h_x = h_z = 0.2$. In each panel, each group of lines is colored with respect to the distance from the boundary, with the top group of lines being the closest to the boundary.}
		\label{supp_fig_Coeff_vs_beta}
	\end{figure}

	Then, in Fig.~\ref{supp_fig_Coeff_vs_beta}, we plot the coefficients of three-body (a) and four-body (b) terms as a function of $\beta$, highlighting again the $\beta^{2(d+1)-n}$ behavior.
	
\begin{figure}
		\centering
		\includegraphics[width=\linewidth]{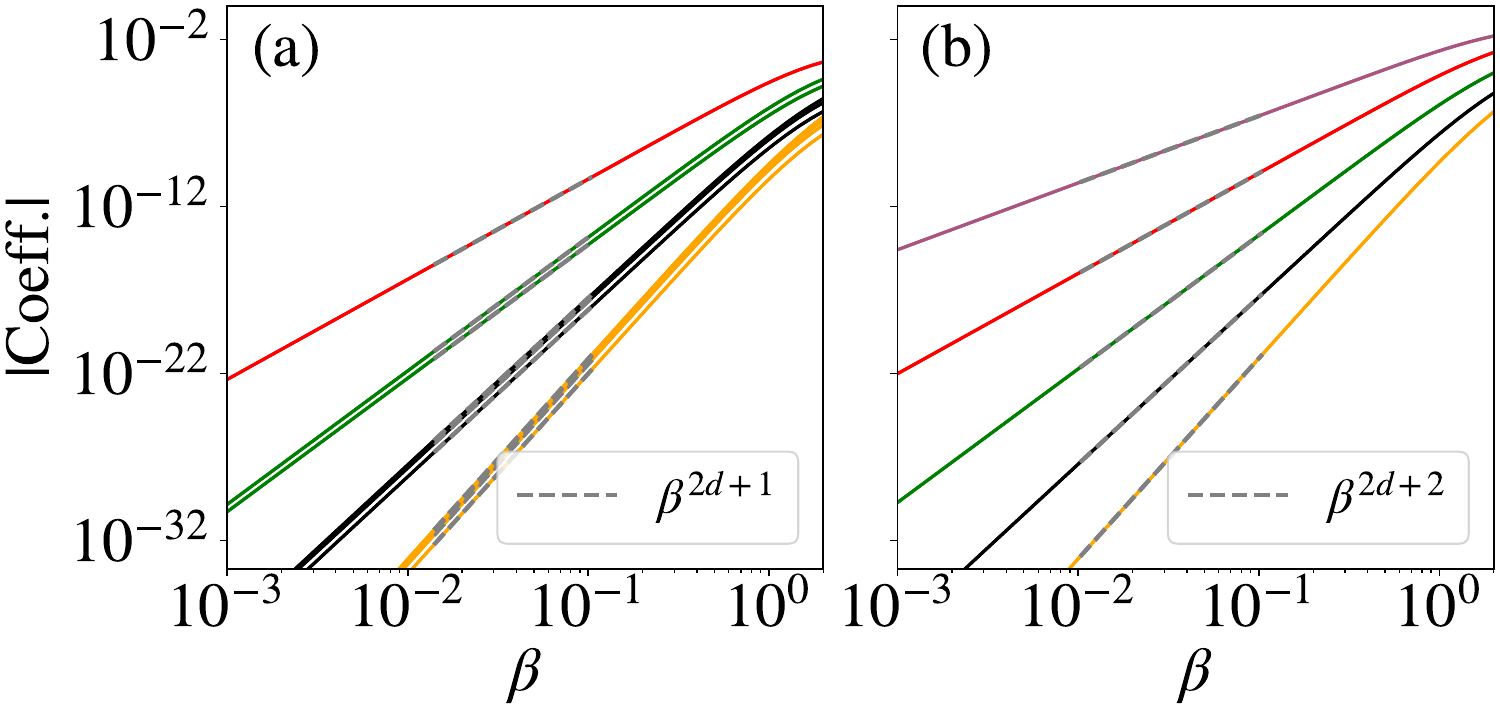}
		\caption{Magnitude of 2-body coefficients, (a) $\sigma_j^{\alpha}\sigma_k^{\alpha'}$ and (b) $\sigma_j^{\alpha}\sigma_{L_A}^{\alpha'}$, in the Hamiltonian of mean force plotted versus $\beta$ - Uniform field \textit{XXZ} chain with $L=7$, $L_A = 6$, $J=1$, $\Delta=1$, $h^x = h^z = 1$. In each panel, each group of lines is colored with respect to the distance from the boundary, with the top group of lines being the closest to the boundary.}
		\label{supp_fig_Coeff_vs_beta_mixed}
	\end{figure}

	Finally, in Fig.~\ref{supp_fig_Coeff_vs_beta_mixed}, we plot the coefficients of the mixed 2-body terms as a function of $\beta$, illustrating that they do not follow equality with the lower bound of $\beta^{2d}$. Fig.~\ref{supp_fig_Coeff_vs_beta_mixed}(a) shows that non-boundary terms scale as $\beta^{2d+1}$, while (b) shows that boundary terms scale as $\beta^{2d+2}$.
	
	\section{Additional data for disordered field model}\label{supp_disorder}
	
\begin{figure}
		\centering
		\includegraphics[width=\linewidth]{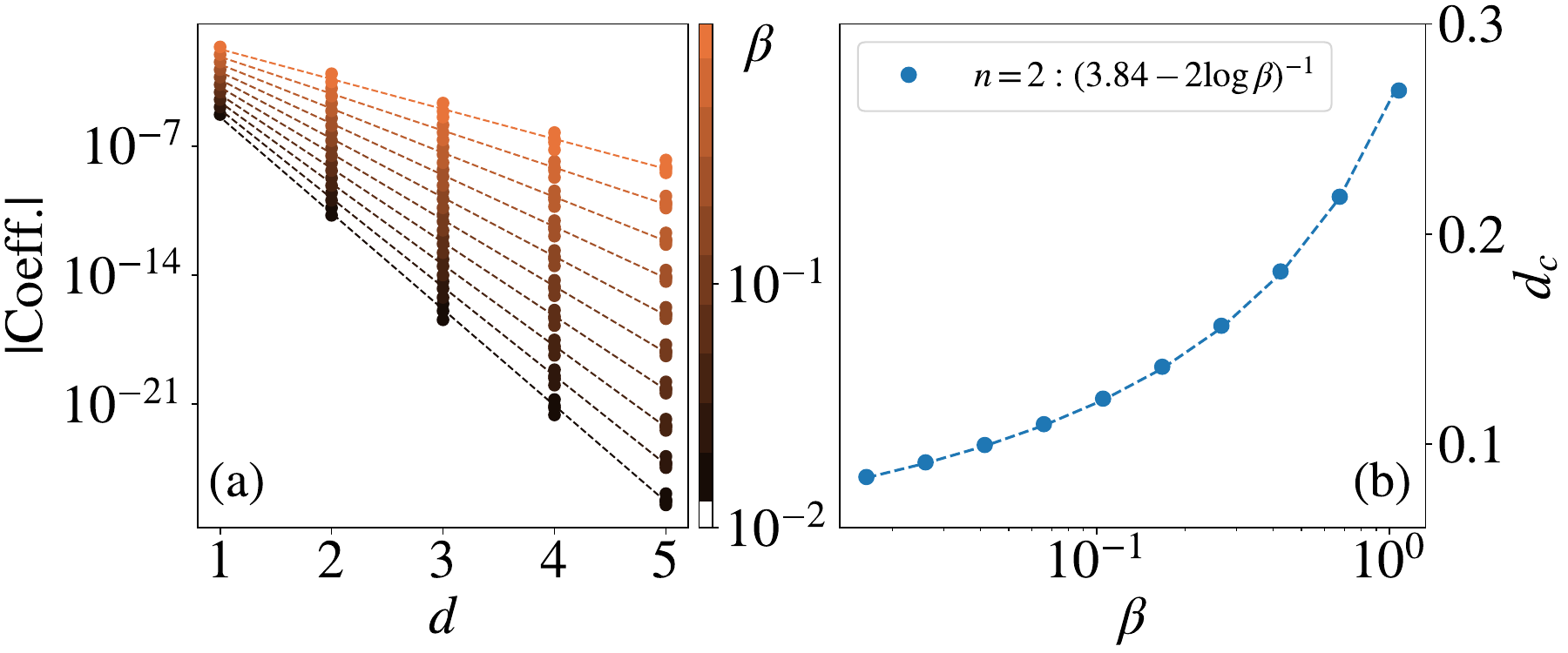}
		\caption{\textbf{(a)} Magnitude of two-body coefficients ($\sigma_j^x \sigma_k^x$) in the HMF versus distance for different $\beta$ (colorbar).
			\textbf{(b)} Skin depth $d_c$ (dots) as a function of $\beta$. $n=1$ corresponds to slopes in (a). Dashed lines are $(a-2\log\beta)^{-1}$ for fitted $a$ (legend).
			Data for a disordered field \textit{XXZ} chain with $L=7$, $L_A=6$, $J=1$, $\Delta=0.95$, and $h_x = h_z = 0.2$.}
		\label{fig:Coeff_vs_d_disorder}
	\end{figure}

\begin{figure}
		\centering
		\includegraphics[width=\linewidth]{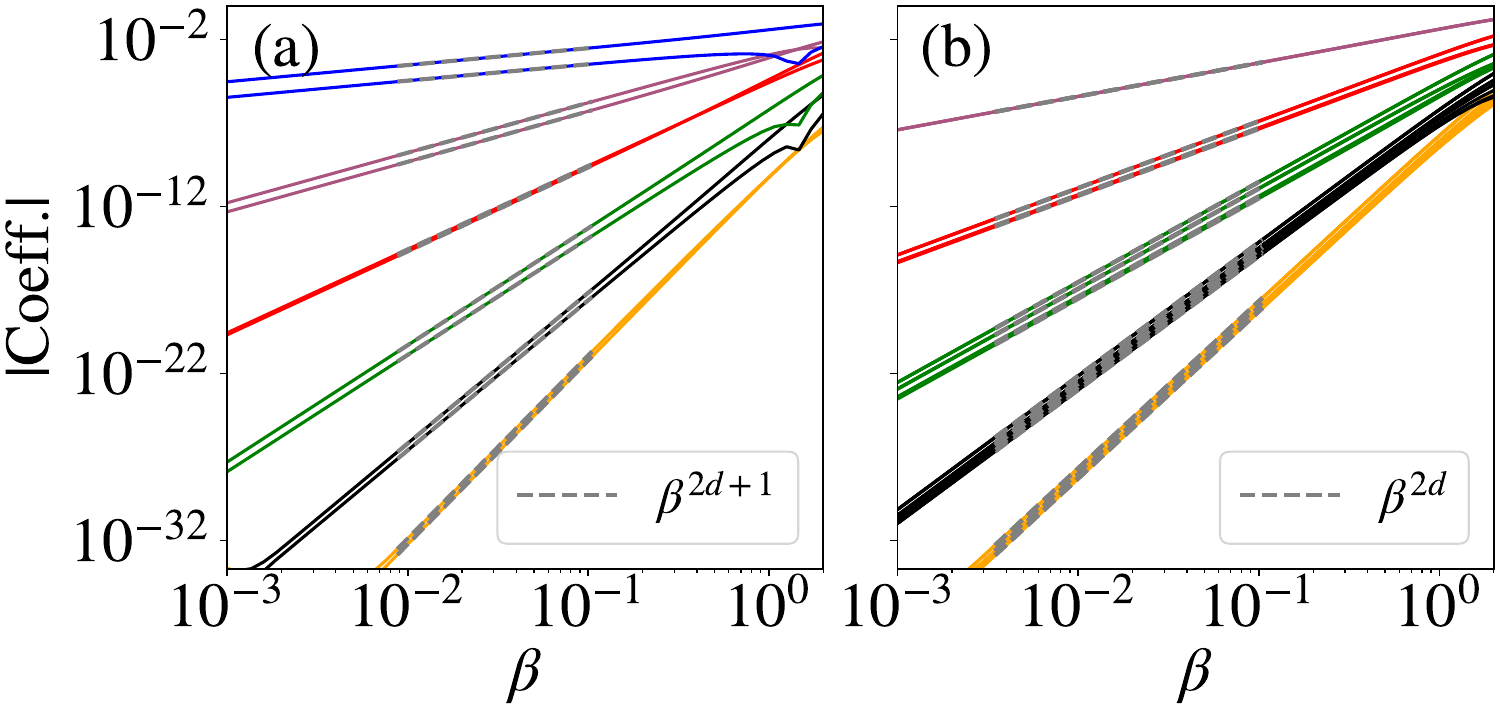}
		\caption{Magnitude of (a) 1-body ($\sigma_j^{\alpha}$) and (b) 2-body ($\sigma_j^{\alpha}\sigma_k^{\alpha}$) coefficients in the Hamiltonian of mean force plotted versus $\beta$ - Disordered field \textit{XXZ} chain with $L=7$, $L_A = 6$, $J=1$, $\Delta=0.95$, $h^x_j, h^z_j \in [-0.2,0.2]$. In each panel, each group of lines is colored with respect to the distance from the boundary, with the top group of lines being the closest to the boundary.}
		\label{fig:Coeff_vs_beta_disorder}
	\end{figure}

	In the main text, we introduced the \textit{XXZ} chain with the addition of transverse and longitudinal magnetic fields,
	\begin{align}\label{eq:disorder_chain}
		H \ = \ J \sum_{j=1}^{L-1}&(\sigma^x_j \sigma^x_{j+1} + \sigma^y_j \sigma^y_{j+1}) + \Delta \sum_{j=1}^{L-1} \sigma^z_j \sigma^z_{j+1} \notag \\
		& +\sum_{j=1}^{L} (h_j^z \sigma^z_j + h_j^x \sigma^x_j).
	\end{align}
	We generally presented data for the case of uniform magnetic field strength across the lattice, however, we have checked our results for various parameters. 
	In Figures \ref{fig:Coeff_vs_d_disorder}-\ref{fig:Coeff_vs_beta_disorder}, we present results using Eq.~\eqref{eq:disorder_chain} with \textit{disordered} magnetic fields, $h^z_j \in [-h^z, h^z], \ h^x_j \in [-h^x, h^x] $. 
	In Fig.~\ref{fig:Coeff_vs_d_disorder} (a), we plot the coefficients of one-body terms versus the distance $d$ to the boundary for various $\beta$ values. In (b), we plot $d_c$ as a function of $\beta$, illustrating the $a-2\log\beta$ behavior.
	In Fig.~\ref{fig:Coeff_vs_beta_disorder}, we plot the coefficients of one and two-body terms as a function of $\beta$.

	\section{Tuning Coupling Strength}\label{supp_tuningJab}
	
\begin{figure}[htb]
		\centering
		\vspace*{0.5cm}
		\includegraphics[width=\linewidth]{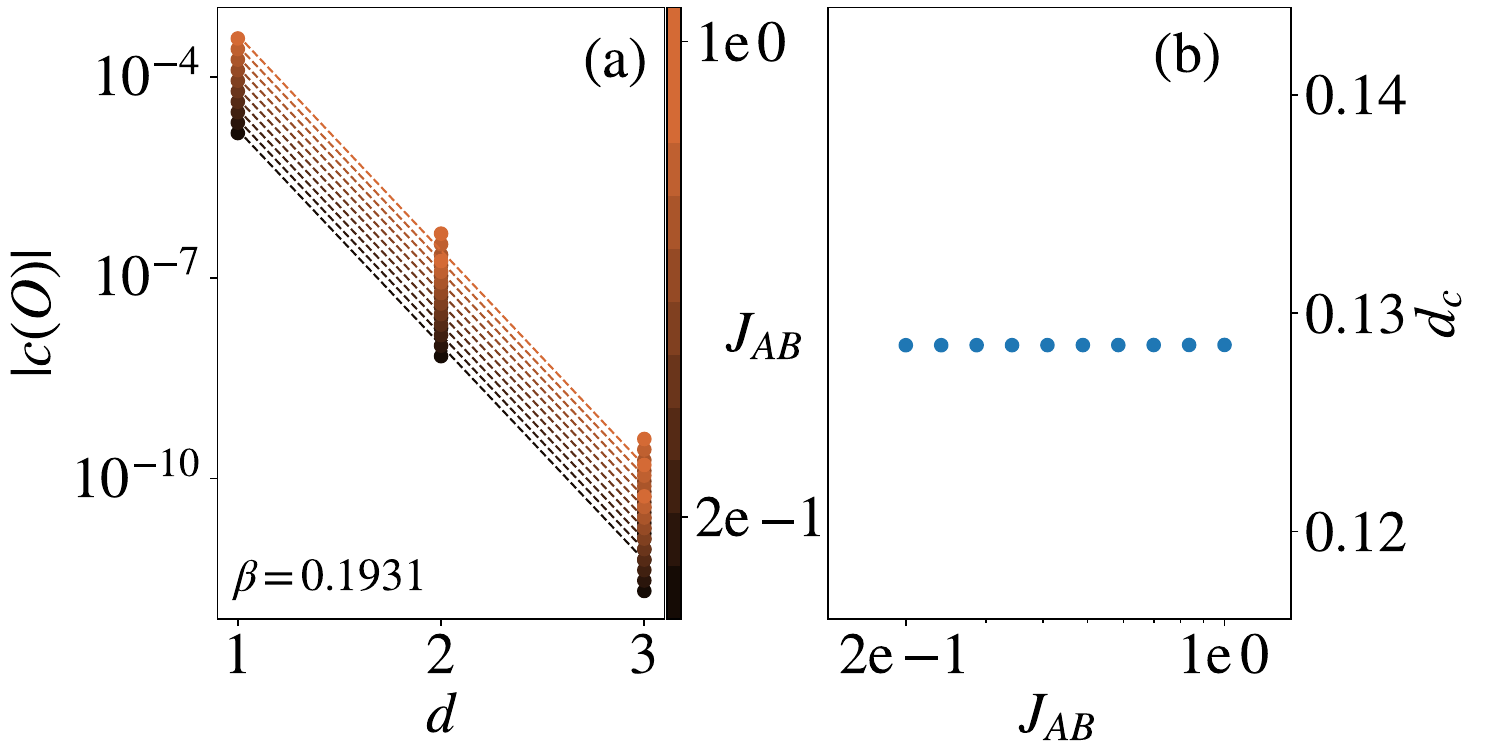}
		\caption{
			\textbf{(a)} $|c(O)|$ for 2-body terms ($\sigma_j^x \sigma_k^x$) in the HMF versus distance $d$ for various $J_{AB}$ (colorbar), at fixed $\beta=0.1931$.
			\textbf{(b)} Skin depth $d_c$ as a function $J_{AB}$ for the same 2-body terms and temperature as in (a). Data for a uniform field \textit{XXZ} Chain with $J = 1$,  $\Delta = 0.95$, $h_x = h_z = 0.2$, $L=6$, $L_A = 4$.
		}
		\label{fig:hmf_coeffsVsd_tuningJab}
	\end{figure}

	In the main text, we stated that the strength of the interaction, $J_{AB}$, does not affect our presented results. In particular, while the magnitude of the coefficients will change, the value of $d_c$ does not change with $J_{AB}$. In Fig.~\ref{fig:hmf_coeffsVsd_tuningJab} (a), we show the value of two-body coefficients as a function of distance $d$ for various $J_{AB}$ at $\beta = 0.1672$. One can see from the figure that the lines shift downwards but do not perceptibly change in slope. In (b), the value of $d_c$ is plotted versus $J_{AB}$, again showing the value hardly changes with decreasing $J_{AB}$. 
	
	\section{Tuning Bath Size}\label{supp_tuningBath}
	
\begin{figure}[b]
		\includegraphics[width=\linewidth]{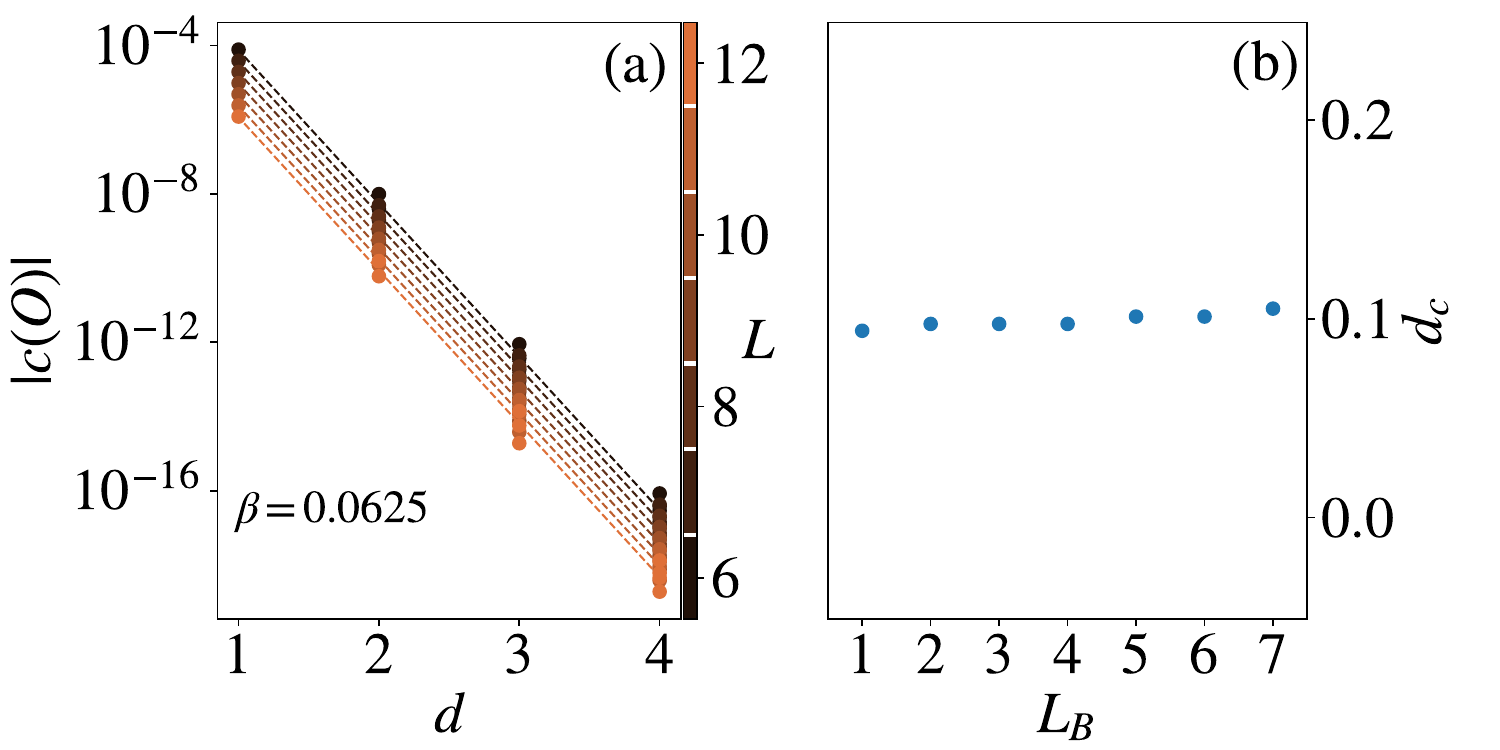}
		\caption{
			\textbf{(a)} $|c(O)|$ for 2-body terms ($\sigma_j^x \sigma_k^x$) in the HMF versus distance $d$ for various $L$ (colorbar), at fixed $\beta=0.0625$.
			\textbf{(b)} Skin depth $d_c$ as a function $L_B (L)$ for the same 2-body terms and temperature as in (a). 
			Data for a uniform field \textit{XXZ} Chain with $J = 1$,  $\Delta = 0.95$, $h_x = h_z = 0.2$, $L_A = 5$.
		}
		\label{fig_hmf_coeffsVsd_tuningLb}
	\end{figure}

	In the main text, we stated that the size of the bath, $L_{B}$, does not affect the exponential behavior detailed therein. In particular, while the magnitude of the coefficients may change, the value of $d_c$ does not meaningfully change with $L_{B}$. In Fig.~\ref{fig_hmf_coeffsVsd_tuningLb} (a), we show the value of two-body coefficients as a function of distance $d$ for various $L_{B}$ at $\beta = 0.0625$. One can see from the figure that the lines shift downwards but do not perceptibly change in slope. 
	In (b), the value of $d_c$ extracted from the lines in (a) are plotted versus $L_B = L - L_A$. We observe that $d_c$ is essentially constant with increasing $L_B$. The minor fluctuations can be attributed to errors in the fitting of the lines in (a), however extracting explicit error bars from the data is cumbersome.

\end{document}